# Exciton ground state fine structure and excited states landscape in layered halide perovskites from combined BSE simulations and symmetry analysis

*Claudio Quarti,[1,]* Giacomo Giorgi,[2] Claudine Katan, [3] Jacky Even,[4,]* Maurizia Palummo[5,]**

[1] Laboratory for Chemistry of Novel Materials, Materials Research Institute, University of Mons, Place du Parc 20, 7000-Mons, Belgium

[2] Department of Civil & Environmental Engineering (DICA), University of Perugia, Via G. Duranti, 93, 06125 Perugia, Italy; CNR-SCITEC I-06123 Perugia, Italy; CIRIAF – Interuniversity Research Centre, University of Perugia, Via G. Duranti 93, 06125 Perugia, Italy

[3] Univ Rennes, ENSCR, INSA Rennes, CNRS, ISCR (Institut des Sciences Chimiques de Rennes) - UMR 6226, F-35000 Rennes, France

[4] Univ Rennes, INSA Rennes, CNRS, Institut FOTON - UMR 6082, F-35000 Rennes, France

[5] Dipartimento di Fisica and INFN, Universitá di Roma "Tor Vergata", Via della Ricerca Scientifica 1, 00133 Roma, Italy

**ABSTRACT**

Layered halide perovskites are solution-processed natural heterostructures where quantum and dielectric confinement effects down to the nanoscale strongly influence the optical properties, leading to stabilization of bound excitons. Achieving a detailed understanding of the exciton properties is crucial to boost the exploitation of these materials in energy conversion and light emission applications, with current on-going debate related to the energy order of the four components of the most stable exciton. To provide theoretical feedback and solve among contrasting literature reports, we perform here ab-initio solution of the Bethe Salpeter Equation (BSE), with detailed interpretation of the spectroscopic observables based on symmetry-analysis. We confirm the $E_{dark} < E_{in\text{-}plane} < E_{out\text{-}of\text{-}plane}$ fine-structure assignment, as from recent magneto-absorption experiments. We further suggest that polar distortions may lead to stabilization of the in-plane component and ultimately end-up in a bright lowest exciton component. Also, we discuss the exciton landscape over a broad energy range and clarify the exciton spin-character, when large spin-orbit coupling is in play, to rationalize the potential of halide perovskites as triplet sensitizers in combination with organic dyes. In addition to contributing to the current understanding of the exciton properties of layered halide perovskites, the work further evidence the in-depth knowledge gained by combining advanced ab-initio simulations and group theory.

Two-dimensional (2D) layered lead-halide perovskites are solution-processed inorganic/organic quantum-well structures, holding great potential for opto-electronic applications. Light emitting diodes based on these materials have set remarkable external quantum efficiencies,[1,2] up to 16.4%,[3] while quasi-2D nanostructures recently demonstrated fair performances,[4] also for the case of traditionally challenging blue-emission.[5,6] In parallel, layered halide perovskites and related nanostructures are intensively scrutinized also for photovoltaics, with recent reports of pure phase 2D compounds rapidly approaching 20% PV-efficiency,[7–9] filling hence the gap with their 3D analogues, meanwhile healing the well-known stability issue of these latter.[10–12] At last, recent reports also demonstrate the potential of single crystal layered halide perovskites for lasing.[13] Overall, these solution-based multi-layered semiconducting superlattices offer unique opportunities to chemically engineer quantum and dielectric confinement effects.[14,15]

In layered 2D lead-halide perovskites, inorganic sheets composed by corner-shared $PbX_6$ octahedra (X=halide) are intercalated by electronically inert, bulky organic cations/dications (A'/A") acting as spacers, the final stoichiometry being $A'_2A_{n-1}Pb_nX_{3n+1}$ or $A''A_{n-1}Pb_nX_{3n+1}$ (A corresponding to a small organic molecule or an alkali metal), with *n* indicating the number of stacked $PbX_6$ layers composing the individual inorganic sheet. The confinement of the photogenerated charges in atomically thin $Pb_nX_{3n+1}$ quantum-wells then imparts these materials with unique optical properties compared to parental 3D halide perovskite compounds, with the enhancement of many-body electronic effects and optical non-linearities.[16,17] Quantum confinement leads in particular to the formation of room temperature stable electron-hole bounded pairs,[18] or excitons, in similar fashion as in transition metal dichalcogenides,[19–22] graphene-derivatives,[23,24] and hexagonal boron-nitride.[25] The dielectric contrast between the inorganic quantum-well and the organic barrier also leads to a deviation of the observed optical signatures from conventional Wannier exciton model in 2D systems,[26,27] further complicating the quest towards a sounded understanding of the optical properties of layered 2D halide perovskites. Nonetheless, pioneering studies before the "perovskite-fever" era provided fundamental contributions in this direction. Namely, early spectroscopic investigations of layered 2D halide perovskites provided estimates of the exciton binding energies ranging from 170 meV,[28] to 370 meV,[26] as function of the organic spacer and of the inorganic quantum-well thickness n, in line with more recent estimates of 450 meV for $(C_4H_9NH_3)_2PbI_4$.[29] In 2005, Tanaka and co-workers successfully identified the 1s-, 2s-, 2p- and 3p- Wannier exciton components for $(C_6H_{13}NH_3)_2PbI_4$ 2D halide perovskite, via combination of UV-vis, two-photon absorption and electro-absorption measurements.[30] Similar understanding for butylammonium lead iodide multi-layered quantum wells was achieved only recently.[29] Pioneering magneto-reflection measurements provided first estimate of Zeeman splitting and diamagnetic shift for $(C_6H_{13}NH_3)_2PbI_4$,[31] and $(C_{10}H_{21}NH_3)_2PbI_4$,[32] with also first potential identification of exciton ground state fine structure.[33] Back in 2005, a first assignment of the exciton ground state fine structure of $(C_4H_9NH_3)_2PbBr_4$ was made on the basis of polarized PL measurements. The following energetic order for the four components expected from symmetry analysis assuming a $D_{4h}$ point symmetry was proposed: $E_{dark} < E_{out\text{-}of\text{-}plane} < E_{in\text{-}plane}$.[34] Very recently, this assignment has been revisited by Do *et al.* on the basis of similar experiments but using the $(C_6H_5\text{-}(CH_2)_2\text{-}NH_3)_2PbI_4$ perovskite, confirming the dark state as the most stable, but finding reversed emission energies for the in-plane and out-of-plane bright components ($E_{in\text{-}plane} < E_{out\text{-}of\text{-}plane}$).[35] Recent magneto-absorption experiments performed on halide perovskites of different compositions, implemented by Dyksik *et al.*, also support reverse order of bright components ($E_{in\text{-}plane} < E_{out\text{-}of\text{-}plane}$), irrespectively of the halide (iodine or bromine) and of the metal (tin or lead) species composing the 2D perovskite.[36] Moreover, all the above-mentioned experimental studies show that in the absence of magnetic field, the exciton ground state is not perfectly dark, indicating that the assumed $D_{4h}$ point symmetry is probably broken.[17]

These contradicting assignments motivate the need for further investigations, with theoretical atomistic simulations at the forefront, thanks to their complementarity with respect to experiments. In this frame, some of the present authors performed fully ab-initio (parameter-free) many-body electronic structure simulations for the thinnest $n$=1,2 compounds of the $(C_4H_9NH_4)_2(CH_3NH_3)_{n-1}Pb_nI_{3n+1}$ series, successfully reproducing the observed excitonic resonance and the exciton binding energy.[37] Following different strategy, some of the present authors performed semi-empirical solution of the Bethe Salpeter Equations (BSE) for the same 2D halide perovskite series, retrieving an analytical expression for the dependence of the exciton binding energy with respect to the quantum well thickness, incorporating dielectric confinement effects.[29] Moreover, they systematically combined experiments and semi-empirical modelling to investigate the deviations of the Ns- exciton components from the 2D Rydberg series, considering various quantum well thicknesses. Overall, numerical solution of the BSE is a more and more exploited tool to investigate the properties of halide perovskite-related materials.[38–50] Inspired by these successful contributions in the field, we perform here full ab-initio, many-body $G_0W_0$/BSE electronic structure calculations, to clarify the exciton fine structure and higher energy exciton resonances in 2D halide perovskites. To set neat connection with predictions from symmetry-analysis,[15,17,34,51] we adopted symmetrized atomistic models as reference rather than experimental structure from XRD measurements, then accounting for the role of octahedral tilting and polar-distortions via group-subgroup analyses and additional simulations. The present work supports the revision of the exciton ground state fine structure proposed by Do *et al.*,[35] Dyksik *et al.*,[36] and nicely parallels the experimental results for higher energy excitons, bridging well-established theories developed for conventional semiconductors and cutting-edge ab-initio numerical simulations.

**RESULTS AND DISCUSSIONS**

Before showing the results from ab-initio calculations, we first recap the main expectations for the excitonic properties of 2D halide perovskites from the group-theory based, symmetry analysis. These are performed considering double-group (then, including SOC) and in absence of external perturbations, as magnetic fields. Symmetry analysis is performed adopting the minimal $Cs_2PbI_4$ reference model in Figure 1a,[34,51] consisting in just one $PbI_6$ octahedron per cell/layer and presenting tetragonal P4mmm space group symmetry. This model is flexible enough to allow exploring various symmetry breaking in a controlled manner, including the ones related to Rashba-Dresselhaus effects on the electronic band dispersions that might produce an effective exchange energy on the exciton fine structure.[52] The proposed minimal model can also be related to the Dion-Jacobson subclass of 2D perovskites, when the interlayer shift is considered for the classification of 2D perovskites instead of the nature of the cation in the barrier.[15] Corresponding Brillouin zone (BZ) and DFT band structure at PBE level (both without and with SOC) are depicted in Figure 1b and 1c, respectively. The latter shows band gap at M and flat dispersion along the M→A direction that reflects the lack of interplanar electronic communication.[53] Symmetry analysis from ref.[51] explains the band gap at M as due to the peculiar Pb(s)-I(p) and Pb(p)-I(s) hybridization for the valence/conduction band edge (VBE/CBE) in this point of the BZ, as compared to Γ (see Supporting Information). Furthermore, it assigns the simple-group symmetry (no SOC) of the frontier orbitals, $\chi_{VBE}$ and $\chi_{CBE}$, to $M_1^+$ and $M_5^-$, irreducible representations (IR), respectively. Inclusion of SOC via double-group analysis leads to the splitting of the two-fold degenerate ($M_5^-$) CBE into two components belonging to $M_6^-$ and $M_7^-$ IR, the VBE transforming as $M_6^+$. To connect these results to the other important subclass of 2D Ruddlesden-Popper (reference I4mmm space group symmetry), we notice that the cell doubling along the stacking axis shifts the band gap at the X point of the corresponding BZ. In absence of SOC, the doubly degenerate CBE then formally splits, as the factor group ($D_{2h}$) in X lacks IR of dimensions larger than one.[15] However when the SOC is included, a doubly degenerated spin-orbit split-off band appears at the CBE in a very similar way than for the minimal $Cs_2PbI_4$ Dion-Jacobson model considered in this work (figure 1c).

Shifting our attention to excitons, these are often described in conventional semiconductors within the frame of the Wannier exciton empirical model. This reduces the problem of a bounded electron-hole pair relative motion to a hydrogenic-like one, whose solutions correspond to 1s-, 2s-, 2p-, etc. resonances, s- or p- symmetry reflecting the Coulomb interaction between the hole and the electron. IR associated to the exciton ($\chi_{exc}$) involving a hole in VBE and an electron in CBE can be therefore easily obtained from:[54]

$$\chi_{exc} = \chi_{VBE}^{*} \otimes \chi_{CBE} \otimes \chi_{env}$$

cenv referring to the IR of the envelope function associated to the relative motion of the electron-hole pair. With exciton ground state associated to a total-symmetric, s-like IR ($\chi_{env}=A_{1g}$), $\chi_{exc}$ corresponds then to $A_{1u}+A_{2u}+E_u$. The exciton fine structure for this minimal model has then four components:

- one state ($A_{1u}$) whose IR is not included in the vectorial representation $A_{2u}$+$E_u$ of the dipole operator, being inaccessible via linear optical excitation (dark);

- one state ($A_{2u}$) with the same IR than the electric dipole operator orthogonal to the perovskite layer, then bright to linear optical measurements with light polarized along the stacking axis;

- two degenerate states ($E_u$) with the same IR of in-plane electric dipoles, also optically bright (degeneracy of the two in-plane components clearly follows the tetragonal symmetry, that is, equivalent a and b in-plane lattice parameters).

The same exciton fine structure holds also for higher energy 2s-, 3s-, etc. (here on, Ns-) components of the Wannier series, although their intensity is expected to decrease with increasing quantum number N.[54] The IR of the 2p-, 3p- etc. components are also easily obtained considering $\chi_{env}$=$E_u$ or $A_{2u}$, leading to $2_{Eg}$, $A_{1g}$, $A_{2g}$, $B_{1g}$, $B_{2g}$ (for $\chi_{env}$=$E_u$) and $E_g$, $A_{1g}$, $A_{2g}$ (for $\chi_{env}$=$A_{2u}$). All p-like states are inaccessible (dark) to linear optical absorption techniques and their identification then requires more involved experimental measurements, like two-photon absorption.[30]

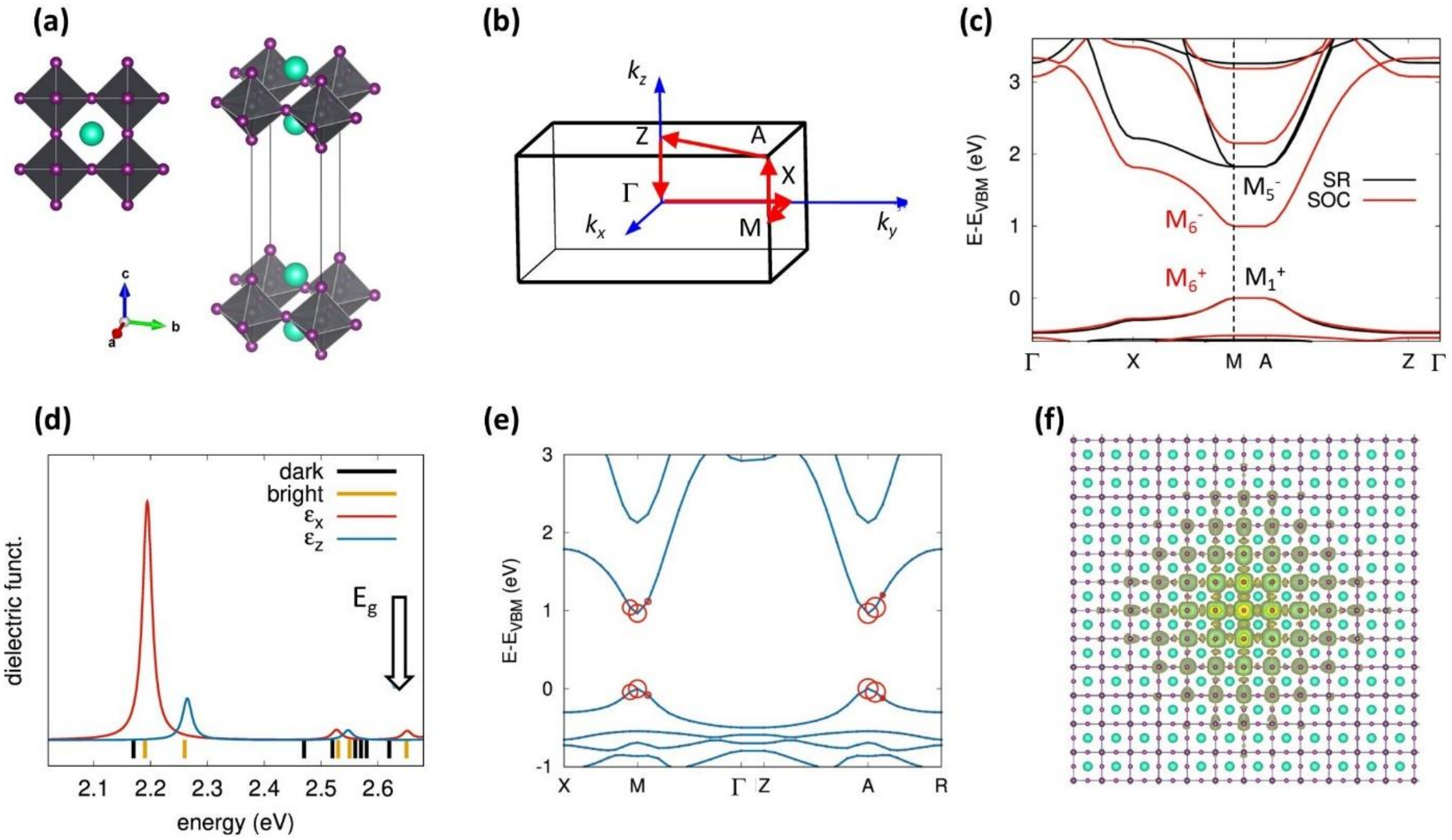

**Figure 1.** Electronic and exciton properties computed for the reference "minimal" 2D halide perovskite model; (a) structural model; (b) corresponding Brillouin zone; c) DFT (PBE) band structure at scalar relativistic (SR) and full-relativistic spin-orbit coupling (SOC) level, with indication of the irreducible representations (IR) of the frontier orbitals; (d) imaginary part of the dielectric function within the inorganic plane ($\varepsilon_x$) and along the stacking direction ($\varepsilon_z$), predicted from $G_0W_0$/BSE simulation. Vertical bars below the dielectric function indicate the excitonic resonances from BSE, with distinction between dark and bright ones. The arrow indicates the $G_0W_0$ corrected band gap ($E_g$); (e) contributions from single particle states in the Brillouin zone for the first dipole allowed $E_u$ transition (e) and corresponding exciton wavefunction (f). This latter shows the first bright exciton wavefunction square modulus (this represents the probability of finding the electron when the hole position is fixed close to a Pb ionic site. The isosurface is set to $10^{-6}$ e/Bohr$^3$).

Given the predictions based on the $D_{4h}$ symmetry analysis, ab-initio simulations are implemented to inspect the ordering of in-plane versus out-of-plane components.[34–36] We start with the iodide-based composition, as it is more relevant for technologic applications due to its red-shifted emission, obtaining single particle band structure from DFT simulations including SOC, then improving the description of the electronic screening via $G_0W_0$ method. Computed band gap amounts to 2.63 eV, consistent with the value reported by Filip *et al.*[46] and in reasonable agreement with experimental measurements of 2.7 eV for $(C_4H_9NH_3)_2PbI_4$.[26,55] Still, further band gap opening is expected when performing self-consistent GW. In addition, we cannot rule out possible error compensation related to the minimal model, namely different self-energy corrections[14,46] stemming from lack of octahedral tilting[38] and incorrect dielectric screening (Cs instead of the organic spacer). Finally, we performed ab-initio solution of the BSE to calculate the excitonic resonances (for details see Methods Section). The computed imaginary part of the dielectric function of $Cs_2PbI_4$ (minimal model) is depicted both for in-plane and out-of-plane directions in Figure 1d. All excitonic resonances are indicated by vertical bars to highlight the presence of dark excited states. Our calculations predict four BSE solutions on the low energy side of the spectrum, a dark exciton at 2.17 eV, two bright degenerate excitons at 2.19 eV with in-plane polarization and another exciton at 2.26 eV with out-of-plane polarization, consistently with symmetry analysis. Therefore, BSE calculations predict degenerate in-plane Eu components as more stable than the out-of-plane $A_{2u}$ one, supporting the recent assignment ($E_{dark} < E_{in\text{-}plane} < E_{out\text{-}of\text{-}plane}$) by Do *et al.*,[35] and Dyksik *et al.*[36] Analysis of the in-plane polarized $E_u$ BSE solution in Figure 1e highlights that this optical transition involves mainly VBE and CBE single particle states. In particular, it is mainly associated to vertical (q=0) excitation at the high symmetry M and A points, with smaller contributions from vertical transitions within 0.026 $Bohr^{-1}$ around these points. Corresponding excitonic wavefunction is illustrated in Figure 1f. It can be analyzed within the Wannier framework, with a S-shape of the envelope as expected. Although not being the focus of the present work, as from the use of a simplified structure (vide supra), the present calculations provide estimates for exciton binding energy (440 meV) and dark-bright splitting (20 meV) in qualitative agreement with reported experimental data on exciton binding energies (450 meV[29] for butylammonium based 2D RP with iodine, or 265/450 meV for phenylethylammonium based 2D RP compounds with iodine/bromine[36]) and dark-bright splitting (22-28 meV[36]), supporting further the reliability of the assignment of the whole ground exciton fine structure. Moreover the differences between iodine and bromine based 2D perovskites related to dielectric effects is qualitatively well reproduced (Figure 2).

The choice of the halide seems also to affect the exciton fine structure of the 2D perovskite lattice, hence justifying additional analyses. On the one hand, Dyksik *et al.* measured increased dark-bright splitting from 22 meV to 28 meV when going from lead-iodide to lead bromine lattice, as shown in Figure 2a, explaining this trend on the basis of stronger electron-hole Coulomb interaction leading to smaller exciton Bohr radius and enhanced exchange interaction.[34,36] On the other hand, the assignment proposed by Takana *et al.* ($E_{out\text{-}of\text{-}plane} < E_{in\text{-}plane}$) is based on PL measurements performed on lead bromine lattice,[34] suggesting that the impact of the halide is not just quantitative but also qualitative. We therefore performed the same $G_0W_0$/BSE simulations on $Cs_2PbBr_4$ lattice and found that the in-plane ($E_u$) component is always more stable than the out-of-plane ($A_{2u}$), irrespectively from the halide, hence definitely disproving the fine structure assignment from Tanaka *et al.*[34] In parallel, we retrieve very similar trend for the dark-bright splitting as function of the exciton binding energy, see Figure 2a, in line with the discussion by Dyksik *et al.*[36]

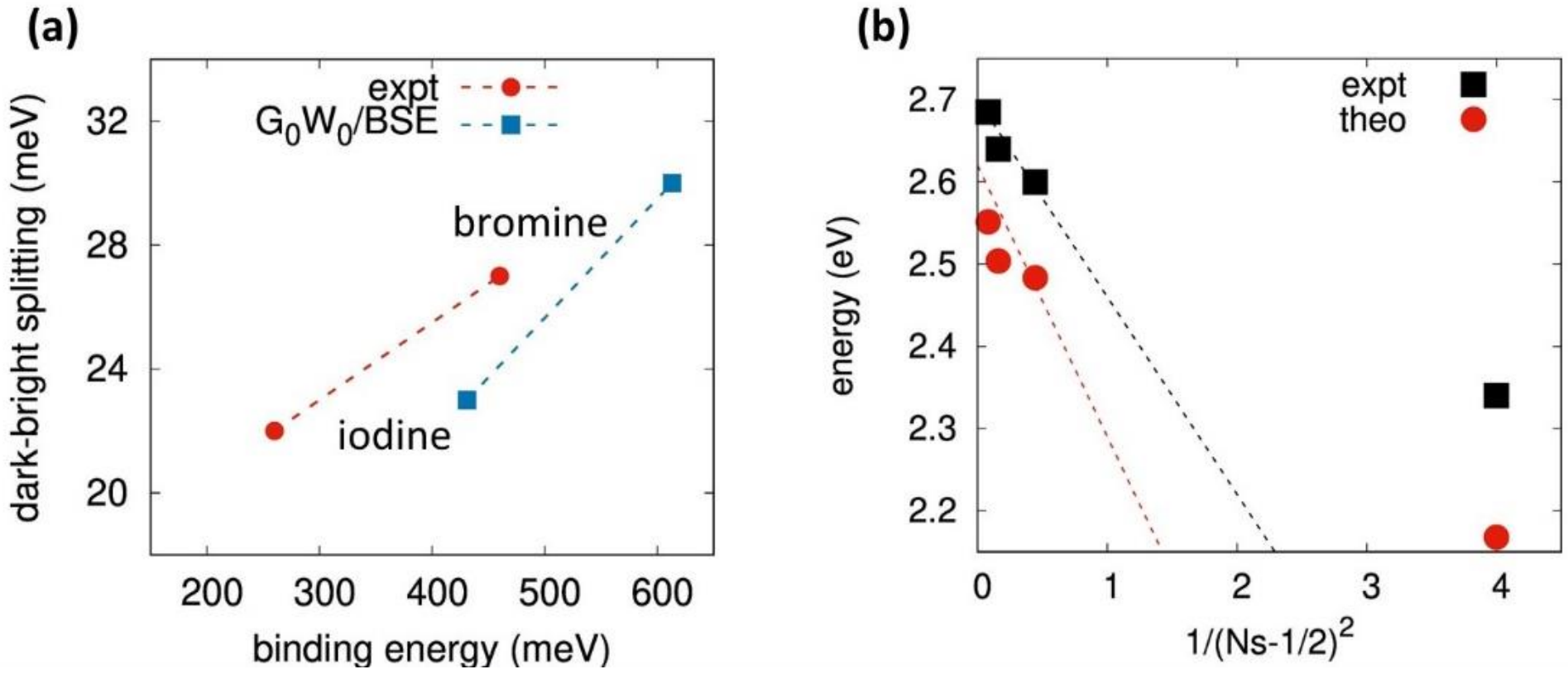

**Figure 2.** (a) dark-bright exciton splitting as function of the exciton binding energy for lead-iodide and lead-bromine lattices. Experimental data refer to phenylethylammonium based 2D RP reported in Ref. [36]; (b) energetics of the Ns- (N=1,4) components of the Wannier-series excitons on $Cs_2PbI_4$ model system and from experimental data from Ref. [26]. The energetics predicted by conventional Wannier model in 2D semiconductors is reported both for experiment and theory, adopting black and red dashed lines, respectively.

Spin-character of excitons in 2D halide perovskites also needs a clarification, as related to the correct usage of singlet and triplet wording for these materials.[56] In absence of SOC, the spin operator commutes with the Hamiltonian and excitons can be therefore classified in terms of pure-spin singlet- and triplet-states. This is the typical case of organic p-conjugated systems, where lack of heavy atoms and corresponding weak SOC leads to the classification of triplet-excitons as optically dark and singlets as bright, at least in the most common case of closed shell spin-singlet ground state. In halide perovskites, the presence of heavy atoms (Pb and Sn) introduces large SOC,[27,53,57] that in turns prevents the system to present pure-spin singlet and triplet states (in present of SOC the Hamiltonian does not commute anymore with the spin operator).[58] First consequence is that the optical activity of the excitons cannot be classified on the basis of the symmetry with respect to pure-spin operator,[56] but rather to the total angular momentum operator *J*=*L*+*S*, as we did above based on double space group symmetry analysis.[51] Second, due to mixed spin-singlet/spin-triplet character of the excited states of halide perovskites, these compounds are currently investigated in combination with spin-pure organic moieties, to take advantage of energy transfer processes at the inorganic/organic interface,[59] with potential exploitation for triplet-sensitization of organic emitters,[60–62] or singlet fission enhancement.[63] The emerging question in this frame is therefore to estimate the "fraction" of spin-singlet and spin-triplet character in these materials. The decomposition of the 1s-exciton fine structure components $A_{1u}$, $E_u$ and $A_{2u}$ from our BSE calculation on pure-singlet ($|S=0, M_S=0\rangle$) and pure-triplet ($|S=1, M_S=0\rangle, |S=1, M_S=\pm1\rangle$) states is reported in Table 1. The weights of the full ab-initio BSE solutions are similarly spread on the various spin states because the Wannier exciton wavefunction is obtained by summation over various vertical transitions in the BZ away from the Γ point. This analysis then indicates large spin-triplet character for the in-plane bright $E_u$ components of the exciton fine structure, mainly due degeneracy, hence explaining the positive performances of halide perovskites as triplet sensitizers, when combined with organic emitters.[60–62] The same decomposition can be performed also considering the simplified Frenkel exciton model developed in Ref. [34]. This incorporates the effect of SOC and the crystal field splitting due to the $D_{4h}$ symmetry, via the effective $\theta$ parameter, with the wavefunction of the four components of the 1s-exciton that are proper eigenstates of the total momentum operator $J$ and $M_J$ ($A_{1u}$, $A_{2u}$ and $E_u$ corresponding to $\left|J=0, M_J=0\right\rangle$, $\left|J=1, M_J=0\right\rangle$, $\left|J=1, M_J=\pm1\right\rangle$ states, respectively). Spin-decomposition for this model is also reported in Table 1, as a function of $\theta$. It highlights again that the fine-structure components shares large contributions from various pure spin-states, although it is more discriminating, compared to the ab-initio results. For instance, it predicts that the in-plane polarized $E_u$ states share largest contribution from one of the two spin-triplet $|S=1, M_S=\pm 1\rangle$ states, at time. Discrepancies of the Frenkel model from Ref.[34] with respect to the ab-initio results should be considered in light of the simplicity of the former, namely the lack of electronic dispersion and the simplified treatment of the tetragonal symmetry, via crystal-field effect.

**Table 1.** Projection of the four exciton-fine structure components on pure states $|S, M_S\rangle$ of the spin-operator $S$. The decomposition is based both on a simplified Frenkel model of a $D_{4h}$ exciton fine structure yielding 4 states classified according to the total momentum $|J, M_J\rangle$ ,[34,64] and on the numerical implementation of the BSE yielding 4 states related to the IR of the $D_{4h}$ point group ($A_{1u}$, $E_u$, $A_{2u}$). The $\theta$ angle is a parameter describing the interplay between the spin-orbit coupling and an effective tetragonal distortion related to the 2D character of the perovskite.[34]

| $\langle J, M_J\|$ \ $\|S, M_S\rangle$ | singlet | triplet | | |
|---|---|---|---|---|
| | $\|0,0\rangle$ | $\|1,0\rangle$ | $\|1,1\rangle$ | $\|1,-1\rangle$ |
| $\langle 0,0\|$ /$A_{1u}$ (dark) | $\frac{cos^2\theta}{2}$/22% | $\frac{cos^2\theta}{2}$/23% | $\frac{sin^2\theta}{2}$/27% | $\frac{sin^2\theta}{2}$/27% |
| $\langle 1,1\|$/$E_{u,1}$ (in plane) | $\frac{sin^2\theta}{2}$/29% | $\frac{sin^2\theta}{2}$/30% | $cos^2\theta$ /21% | 0%/19% |
| $\langle 1,-1\|$/$E_{u,2}$ (in plane) | $\frac{sin^2\theta}{2}$/29% | $\frac{sin^2\theta}{2}$/30% | 0%/19% | $cos^2\theta$ /21% |
| $\langle 1,0\|$/$A_{2u}$ (out of plane) | $\frac{cos^2\theta}{2}$/19% | $\frac{cos^2\theta}{2}$/16% | $\frac{sin^2\theta}{2}$/32% | $\frac{sin^2\theta}{2}$/32% |

Understanding of the exciton manyfold over a broad energy range is important to understand hot-electron relaxation mechanisms. On the other hand, convergence of ab-initio BSE calculations with respect to the sampling of the reciprocal lattice is usually very challenging,[20,21] and this becomes even more problematic when dealing with high energy excited states. We therefore repeated our BSE calculation with denser k-points mesh, finding very small effect on the BSE solutions (see Supporting Information). Further extrapolation of the energetics of the bright BSE solutions for an infinitely dense k-point sampling, as explained in the Supporting information, suggests a stabilization of the exciton ground-state fine structure by ca. 20-50 meV, but preserving energy order $E_{dark}(A_{1u}) < E_{in\text{-}plane}(E_u) < E_{out\text{-}of\text{-}plane}(A_{2u})$.[35,36] Higher energy BSE solutions are more sensitive to the k-point sampling, as expected,[20] with impact on the order of 150 meV. Within these corrections at hand, we proceeded to assign the higher lying solutions from our BSE simulations. Symmetry analysis greatly ease the interpretation, as it anticipates that higher lying Ns- (N>1) excitons of the Wannier series have bright components, in striking contrast to Np-excitons. This allows us to assign the 2s-, 3s- and 4s- components from our BSE calculations, based on the computed dipole moment (lowest 44 BSE solutions listed in Supporting Information). Estimated energies for the Ns- Wannier excitons are illustrated in Figure 2b, along with the experimental data from Ref.[26]. Our BSE simulations nicely parallel the experiment and correctly predict the deviation of the Ns- energetic progression predicted from simple 2D Wannier model, that is a widely-established signature of charge image effects.[14,26] Solutions of our BSE simulations also fit the expectation from symmetry analysis for 2p- Wannier components. Above the 2s- signature, we find in fact 12 excited states, all dark and composed by three pairs-of degenerate $E_g$ states. These states can be further classified into 8 components related to in-plane p-like envelope ($\chi_{env}$=$E_u$), out of which 2 are two-fold degenerate, and 4 related to p-like out-of-plane envelope ($\chi_{env}$=$A_{2u}$), out of which 1 is two-fold degenerate. Similar arguments hold for the $3p_{xy}$ and $3p_z$ components. Out of the 44 lowest energy excitations, only four excited states do not fit the expectations from Wannier-exciton series, lying between the 1s- and 2s- components and being dark. For these states, we are not aware of any experimental confirmation, either from linear or two-photon measurements.[26,30,34,55] Apart from this discrepancy, the overall agreement between ab-initio many-body results and the expectations from conventional theory of semiconductors is remarkable.

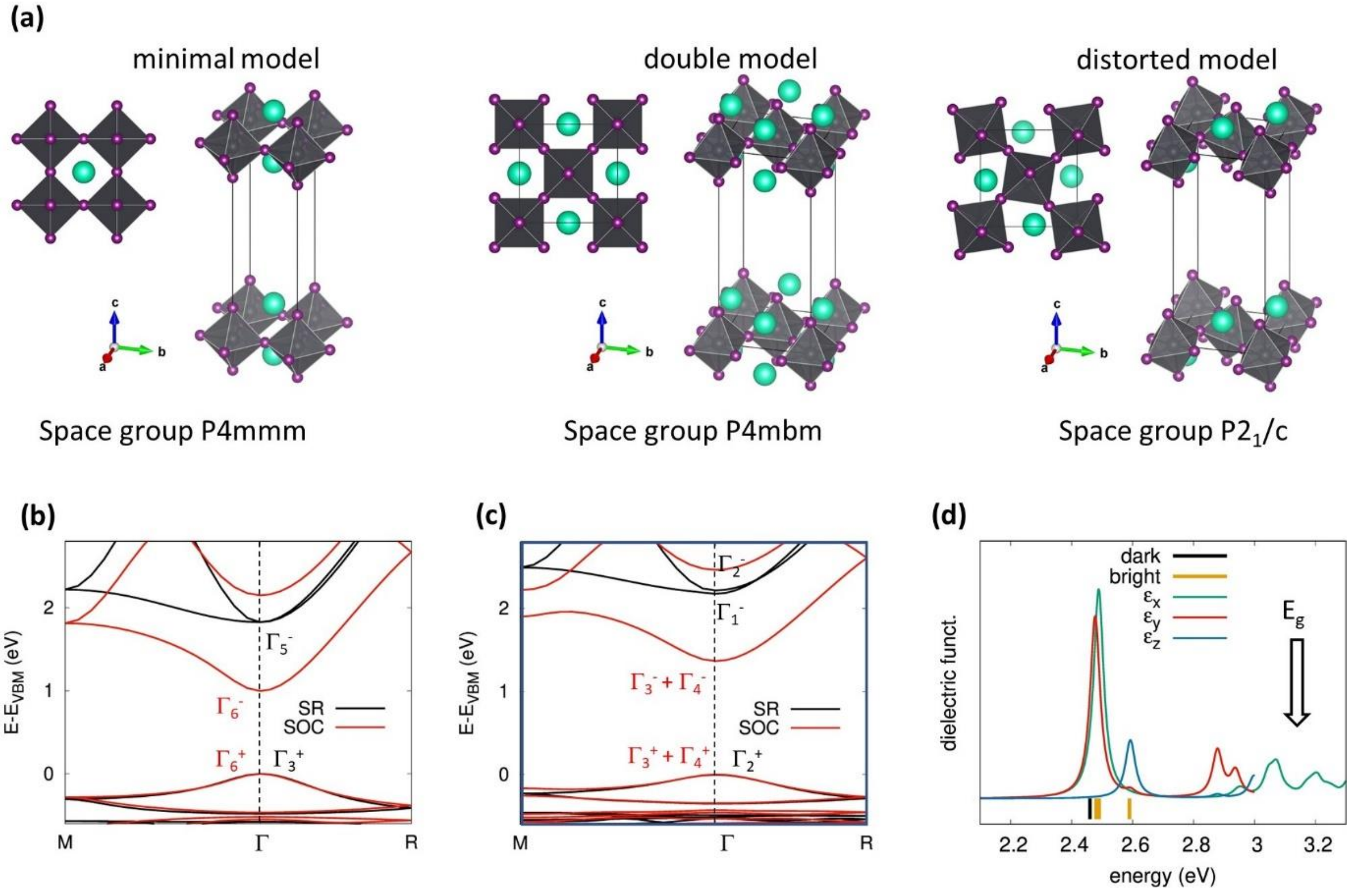

**Figure 3.** Electron and exciton properties in presence of (non-polar) octahedral tilting distortions. (a) incorporation of the octahedral tilting via doubling of the minimal model and rotation of the octahedral along the pseudo-cubic axes; (b) DFT (PBE) band structure for P4mbm "doubled" model and (c) for model including octahedral tilting. Band structures are obtained both at scalar-

relativistic (SR) and full-relativistic (SOC) level, with indication of the IR of the frontier levels. (d) Imaginary part of the dielectric function along the three crystalline axes ($\varepsilon_x$, $\varepsilon_y$, $\varepsilon_z$) for the distorted model, as predicted from $G_0W_0$/BSE simulations. Vertical bars below the dielectric functions indicate the components of the ground state exciton fine structure from BSE, with distinction between dark and bright excitons. The arrow indicates the $G_0W_0$ corrected band gap ($E_g$).

We now move to address the role of detailed structural non-polar and polar distortions. Octahedral tilting represents the most common form of the former type of distortion in both oxide and halide perovskite structures,[65] and they strongly influence the electronic properties of the latter.[38,66,67] Inclusion of antiferrodistorsive non-polar octahedral tilting first requires in-plane cell doubling, as shown in Figure 3a, the intermediate "double" model being therefore equivalent to the "minimal" model from Figure 1a. Group-subgroup analysis assigns this model to P4mbm space group, with IR retrieved from subduction problem for the P4mmm space-group (see Supporting Information),[68,69] VBE and CBE then corresponding to $\Gamma_3^+$ and $\Gamma_5^-$ IR, respectively. The change in the symmetry label from M to Γ simply reflects the band folding of the frontier levels, as shown in the DFT band structure in Figure 3b. Further inclusion of SOC leads to $\Gamma_6^+$ and $\Gamma_6^-$ IR for VBE and CBE, respectively, with exciton ground state fine structure equivalent to the one of the minimal model ($A_{1u}$+$A_{2u}$+$E_u$). Octahedral tilting is then introduced via DFT relaxation of the atomic positions, with final structure in Figure 3a presenting both in-plane and out-of-plane octahedral rotations. Projection of the Pb-I-Pb bond angle on the perovskite plane and angular deviation of the Pb-apical I bond from the normal direction amount to 164 and 7 degrees, respectively. These distortions are accompanied by an increase of the single particle PBE+SOC band gap (1.12 eV) compared to the undistorted model (0.97 eV). Following group-subgroup relation from Supporting Information,[68,69] the final structure presents $P2_1/c$ space group symmetry, monoclinic space groups being indeed often proposed in the refinement of layered perovskite crystal structures.[70–74] Notice that our model distorted cell does not exhibit the traditional shear strain perpendicular to the binary axis characteristic of $P2_1/c$, but it could be easily added since this strain component behaves as the totally symmetric IR $A_g$. The decrease from tetragonal to monoclinic (or similarly orthorhombic) space group leads to two important consequences. First, due to the lack of irreducible representations (IR) of dimension larger than one, the twice degenerate in-plane $E_u$ exciton splits in two components, as nicely found also in our BSE calculations in Figure 3d, the splitting amounting to ca. 10 meV. Similar conclusion actually holds also for the CBE in absence of SOC, that is two-fold degenerated in absence of distortion but that splits into two components $\Gamma_1^-$ and $\Gamma_2^-$ once octahedral tilting is introduced (subduction problem reported in Supporting Information). The second important consequence is that all the components of the exciton fine structure are now optically bright and correspond to $2A_u+2B_u$ IR. Octahedral distortion then provides formerly dark $A_{1u}$ with optical activity, that might explain the systematic experimental observation of a weakly bright exciton ground state for 2D perovskites.[34,36] By contrast a minimal antiferrodistorsion in a 3D lattice is not enough to brighten the exciton dark state.[64]

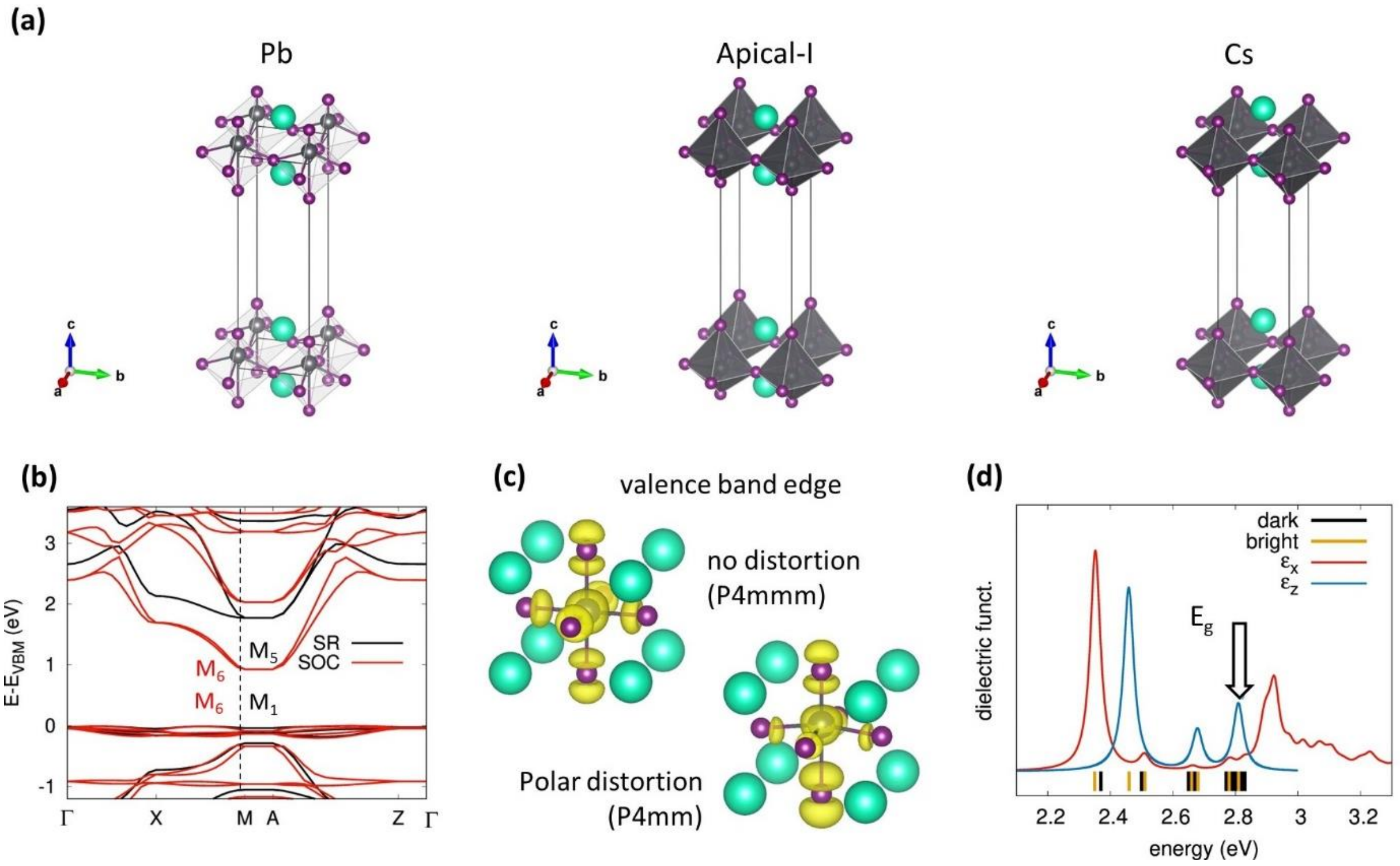

**Figure 4.** Electron and exciton properties in presence of (out-of-plane) polar distortions. (a) polar distortions obtained by displacing either the Pb, apical-I or Cs atom. (b) DFT (PBE) band dispersion for the Pb-distorted model, both at scalar relativistic (SR) and full-relativistic (SOC) level, with indication of the irreducible representations for VBE and CBE. (c) single particle VBE wavefunction square modulus (isosurface set to $2\ 10^{-3}$ e/Bohr$^3$]. (d) imaginary part of the dielectric function in the organic-plane ($\varepsilon_x$) and along the stacking direction ($\varepsilon_z$), computed from $G_0W_0$/BSE calculations. Vertical bars below the dielectric function indicate the excitonic resonances from BSE, with distinction between dark and bright ones. The arrow indicates the $G_0W_0$ corrected band gap ($E_g$).

We close discussing the potential impact of polar distortions on the optical properties of 2D halide perovskites. Becker and co-workers associated the high brightness of 3D $CsPbBr_3$ nanoparticles to the stabilization of the bright triplet exciton fine structure component compared to the dark singlet. In particular, these authors proposed that Rashba splitting, triggered by polar distortions and consequent loss of inversion symmetry, may behave as an effective exchange energy, hence affecting the energy order of the exciton ground-state fine structure component.[52] Subsequent low temperature absorption measurements with magnetic fields performed on $FAPbBr_3$ and $CsPbI_3$ nanoparticles ultimately overturned the proposed inversed stability of the dark/bright excitons, so reducing the potential role played by Rashba splitting and polar distortions, at least in the case of 3D halide perovskites.[75,76] The depolarization field that limits the influence of polar distortions in nanoparticles, in turn, may be less important in 2D perovskites.[77] It is thus interesting to verify "in-silico" whether polar distortions may influence the exciton fine structure of layered halide perovskites. Archetypal ferroelectric distortion in the perovskite structure is associated to displacive atomic motion from high symmetry positions,[78] but in the case of halide perovskites systematic structural descriptors have been introduced, leading to inversion symmetry breaking. Indeed, specific bond angle and in-plane octahedral tilting disparity lead to ferroelectric space group and in turn Rashba-like spinor splitting, as proposed by Jana *et al.*[79] This type of distortions however requires two account different in-plane octahedral rotations, hence imposing large cells, meanwhile leading to poor symmetry. Therefore, in the spirit of performing a rigorous symmetry-based analysis, we consider only simple displacive motions. We then manually displaced either the Pb atom, the apical-I or the Cs atom along the plane stacking direction, by 2% of the stacking lattice vector (see Figure 4a). All these distortions lead to the same P4mm space group, with simple-group analysis anticipating first a different atomic orbital hybridization scheme, compared to the P4mmm minimal model (see Supporting Information), and simple-group subduction problem suggesting $M_1$ and $M_5$ IR for VBE and CBE, respectively (the lack of $\pm$ sign simply reflecting the loss of inversion symmetry). Further inclusion of SOC and analysis of the exciton fine structure leads to $A_1$+$A_2$+E IR, meaning that the exciton ground state of the system presents still one dark state ($A_2$), one state with out-of-plane polarization ($A_1$) and two-fold degenerate states with in-plane polarization (E). Despite these conclusions being general for all the three models in Figure 4a, striking results are found only for the case of the Pb-displacement, as shown in Figure 4b-d. First, single particle band structure is strongly perturbed from the loss of inversion symmetry, with the valence band narrowing and VBE displaced from the high symmetry M point, as shown in Figure 4b. This is in line with the reduced contribution from the equatorial iodines on the VBE compared to the undistorted minimal model, see Figure 4c. In addition, the displacement of Pb from high symmetry position leads to a contribution from the corresponding $p_z$ orbitals, appearing as an asymmetric probability amplitude for an electron to lie in the VBE state, around this atom. Second, BSE simulation computed for this model indeed reveals a reversed ordering of the dark and E-bright states, the latter lying ca. 10 meV at lower energy than the former. The present finding is therefore consistent with the proposition by Becker and takes place in presence of spontaneous polarization along the interplanar stacking of just 1.1 μC/cm$^2$ (see Supporting Information),[52] that is, four times smaller than upper limit of the spontaneous polarization estimated for 3D $CH_3NH_3PbI_3$ halide perovskite.[80] In this sense, inducing polar distortion in layered 2D halide perovskites is in principle a viable strategy to improve the optical properties of these compounds, thanks to the elimination of detrimental bright-to-dark relaxation pathway, the real challenge being then to find suitable ways to practically introduce such distortion in real compounds. Appropriate choice of organic spacer may be a chemically viable route, notwithstanding the recent report of para-substituted benzylammonium lead iodide perovskite, lacking inversion symmetry.[81] In parallel, application of a hydrostatic pressure may also induce polar structural distortions due to steric hindrance and related reverse dark/bright stability, although there are contrasting reports on the influence of pressure on the emission efficiencies of layered halide perovskites.[82,83] Results for polar displacement of apical-I and Cs atoms are reported in the Supporting Information, for sake of completeness. None of them led to similar inversion of Dark and Bright components.

## CONCLUSIONS

To sum up, the present work combines ab-initio many-body simulations and symmetry analysis to provide theoretical support for the comprehension of excitonic properties of 2D lead halide perovskites, addressing in particular the assignment of the exciton ground state fine structure, the identification of higher lying states, the exciton spin-character in presence of large SOC and the role of octahedral tilting and polar distortions. Theoretical results confirm the recent assignment of the exciton fine structure proposed by Do *et al.*,[35] and Dyksik *et al.*,[36] the doubly degenerate in-plane ($E_u$) component being more stable than the out-of-plane ($A_{2u}$) one. Furthermore, we found that the results of ab-initio $G_0W_0$/BSE simulations fit very well the progression expected from simple Wannier-exciton model, with charge image effect implicitly accounted, overall setting a bridge between modern ab-initio numerical approaches and conventional theory from semiconductor physics. The effects of the octahedral tilting are also discussed on the basis of the symmetry reduction from tetragonal to monoclinic space group, leading to splitting of the in-plane degenerate ($E_u$) exciton components by ca. 10 meV, and the formal brightening of the lowest energy dark exciton. Bright-to-dark exchange is finally discussed in presence of a polar distortion of Pb. Overall, the present findings both address important open questions from the literature and provide a general and sounded framework for the discussion of the excited state properties of layered halide perovskites, paving the way for further studies related to the exciton relaxation in broad energy range and energy transfer mechanisms at the interface with organic dyes (for which pure spin-states are relevant) or other low dimensional systems. In parallel, we demonstrate that the combination of advanced ab-initio simulations with detailed symmetry analysis from group-theory has immense potential for in-deep interpretation of numerical results, motivating similar analyses also for other materials of technological relevance.

**Methods**

All the DFT calculations have been performed by means of the Quantum-Espresso suite.[84,85] Norm conserving fully relativistic pseudopotentials to include SOC[86] together with the electronic exchange-correlation functional of Perdew-Burke-Ernzerhof (PBE)[87] and a kinetic energy cut-off of 120 Ry, have been used.

The Yambo code[88,89] is then exploited in order to get the quasi-particle (QP) energies in the $G_0W_0$ one-shot perturbative approach and the optical excitation energies and spectra solving the BSE.[90] The plasmon–pole approximation for the calculation of the inverse dielectric matrix, 10 Ry (70 Ry) has been used for the correlation part of the self-energy $\Sigma c$ (exchange, $\Sigma x$). The same theoretical approach has been recently successfully applied to the study of other 2D, mixed 2D/3D, and 3D hybrid and full-inorganic halide perovskites.[37,39–41] Unoccupied states have been summed up to 25 eV above the valence band maximum. A k-grid of 10x10x2 (20x20x2) is enough to obtain convergence within 0.05 eV for $G_0W_0$ (BSE).

ASSOCIATED CONTENT

**Supporting Information**. Analysis of symmetry adapted linear combinations of atomic orbitals for non distorted and distorted structures. Character tables for all space/point groups investigated (P4mmm, P4mbm, $D_{4h}$, P21/c, $D_{2h}$, P4mm and $D_4$). List of solutions from the $G_0W_0$/BSE simulations performed on the minimal model in Figure 1a. Results from additional $G_0W_0$/BSE calculations performed with denser k-point grid. Subgroup/supergroup relations for the determinations of the space group symmetry of the distorted 2D halide perovskite models. Subduction relationships relating the irreducible representations of the distorted models with those from the minimal one. Berry phase calculation for the estimate of the spontaneous polarization of ferroelectric distorted models. Results from DFT and $G_0W_0$/BSE calculations of ferroelectric distorted I-ap and Cs models. The following files are available free of charge.

AUTHOR INFORMATION

**Corresponding Authors**

*Claudio Quarti: claudio.quarti@umons.ac.be

*Jacky Even: jacky.even@insa-rennes.fr

*Maurizia Palummo: maurizia.palummo@roma2.infn.it

**Author Contributions**

CQ, CK and JE conceived and planned the work. MP and GG performed all the ab-initio calculations. CQ and JE performed symmetry analysis. CQ analyzed the results and prepared the original draft. The manuscript was written through contributions of all authors. All authors have given approval to the final version of the manuscript.

**Funding Sources**

Union’s Horizon 2020, FET Open research and innovation program (grant agreement No 899141 - PoLLoC).

Agence Nationale pour la Recherche (grant No ANR-18-CE05-0026 MORELESS project).

INFN for the TIME2QUEST project.

Tor Vergata University for the TESLA project.

Institut Universitaire de France

FNRS and M-ERA.NET network project (grand No R.8003.22 PHANTASTIC).

ACKNOWLEDGMENT

J.E. acknowledges the financial support from the Institut Universitaire de France. The work at ISCR and Institut FOTON was performed with funding from the European Union’s Horizon 2020 program, through a FET Open research and innovation action under the grant agreement No 899141 (PoLLoC). At ISCR the work was also supported by the Agence Nationale pour la Recherche (MORELESS project). M.P. acknowledges INFN and Tor Vergata University through the TIME2QUEST and TESLA projects respectively. The work at the University of Mons was performed within the frame of the M-ERA.NET project PHANTASTIC (R.8003.22), supported by the FNRS. CQ. is FNRS research associate. JE and CK thank Dr. Edoardo Mosconi and Prof. Filippo De Angelis for early discussions. CQ takes the occasion to thank Prof. David Beljonne for insightful discussions.

ABBREVIATIONS

BSE, Bethe Salpeter Equation; XRD, X-Ray Diffraction; SOC, spin-orbit-coupling; BZ, Brillouin zone; VBE/CBE, Valence/Conduction Band Edge; IR, Irreducible Representations; DJ, Dion Jacobson; RP, Ruddlessden Popper.

# Exciton ground state fine structure and excited states landscape in layered halide perovskites from combined BSE simulations and symmetry analysis

**SUPPORTING INFORMATION**

*Claudio Quarti,[1,]* Giacomo Giorgi,[2] Claudine Katan, [3] Jacky Even,[4,]* Maurizia Palummo[5,]**

[1] Laboratory for Chemistry of Novel Materials, Materials Research Institute, University of Mons, Place du Parc 20, 7000-Mons, Belgium

[2] Department of Civil & Environmental Engineering (DICA), University of Perugia, Via G. Duranti, 93, 06125 Perugia, Italy; CNR-SCITEC I-06123 Perugia, Italy; CIRIAF – Interuniversity Research Centre, University of Perugia, Via G. Duranti 93, 06125 Perugia, Italy

[3] Univ Rennes, ENSCR, INSA Rennes, CNRS, ISCR (Institut des Sciences Chimiques de Rennes) - UMR 6226, F-35000 Rennes, France

[4] Univ Rennes, INSA Rennes, CNRS, Institut FOTON - UMR 6082, F-35000 Rennes, France

[5] Dipartimento di Fisica and INFN, Universitá di Roma "Tor Vergata", Via della Ricerca Scientifica 1, 00133 Roma, Italy

**Table SI1.** Character table of point group $D_{4h}$, of P4mmm space group at the Γ and M points of the corresponding Brillouin zone and of P4mbm space group at Γ.

| **$D_{4h}$** | **P4mmm** | | **P4mbm** | **E** | **$2C_4(z)$** | **$C_2$** | **$2C'_2$** | **$2C''_2$** | **i** | **$S_4$** | **$\sigma_h$** | **$2\sigma_v$** | **$2\sigma_d$** | | |
|---|---|---|---|---|---|---|---|---|---|---|---|---|---|---|---|
| | **Γ** | **M** | **Γ** | | | | | | | | | | | | |
| simple-group | | | | | | | | | | | | | | | |
| $A_{1g}$ | $\Gamma_1^+$ | $M_1^+$ | $\Gamma_1^+$ | 1 | 1 | 1 | 1 | 1 | 1 | 1 | 1 | 1 | 1 | | $x^2+y^2,z^2$ |
| $A_{2g}$ | $\Gamma_3^+$ | $M_3^+$ | $\Gamma_3^+$ | 1 | 1 | 1 | -1 | -1 | 1 | 1 | 1 | -1 | -1 | | |
| $B_{1g}$ | $\Gamma_2^+$ | $M_2^+$ | $\Gamma_2^+$ | 1 | -1 | 1 | 1 | -1 | 1 | -1 | 1 | 1 | -1 | | $x^2-y^2$ |
| $B_{2g}$ | $\Gamma_4^+$ | $M_4^+$ | $\Gamma_4^+$ | 1 | -1 | 1 | -1 | 1 | 1 | -1 | 1 | -1 | 1 | | xy |
| $E_g$ | $\Gamma_5^+$ | $M_5^+$ | $\Gamma_5^+$ | 2 | 0 | -2 | 0 | 0 | 2 | 0 | -2 | 0 | 0 | | xz,yz |
| $A_{1u}$ | $\Gamma_1^-$ | $M_1^-$ | $\Gamma_1^-$ | 1 | 1 | 1 | 1 | 1 | -1 | -1 | -1 | -1 | -1 | | |
| $A_{2u}$ | $\Gamma_3^-$ | $M_3^-$ | $\Gamma_3^-$ | 1 | 1 | 1 | -1 | -1 | -1 | -1 | -1 | 1 | 1 | z | |
| $B_{1u}$ | $\Gamma_2^-$ | $M_2^-$ | $\Gamma_2^-$ | 1 | -1 | 1 | 1 | -1 | -1 | 1 | -1 | -1 | 1 | | |
| $B_{2u}$ | $\Gamma_4^-$ | $M_4^-$ | $\Gamma_4^-$ | 1 | -1 | 1 | -1 | 1 | -1 | 1 | -1 | 1 | -1 | | |
| $E_u$ | $\Gamma_5^-$ | $M_5^-$ | $\Gamma_5^-$ | 2 | 0 | -2 | 0 | 0 | -2 | 0 | 2 | 0 | 0 | x,y | |
| double-group | | | | | | | | | | | | | | | |
| $E_{1/2g}$ | $\Gamma_6^+$ | $M_6^+$ | $\Gamma_6^+$ | 2 | $\sqrt{2}$ | 0 | 0 | 0 | 2 | $\sqrt{2}$ | 0 | 0 | 0 | | |
| $E_{3/2g}$ | $\Gamma_7^+$ | $M_7^+$ | $\Gamma_7^+$ | 2 | $-\sqrt{2}$ | 0 | 0 | 0 | 2 | $-\sqrt{2}$ | 0 | 0 | 0 | | |
| $E_{1/2u}$ | $\Gamma_6^-$ | $M_6^-$ | $\Gamma_6^-$ | 2 | $\sqrt{2}$ | 0 | 0 | 0 | -2 | $-\sqrt{2}$ | 0 | 0 | 0 | | |
| $E_{3/2u}$ | $\Gamma_7^-$ | $M_7^-$ | $\Gamma_7^-$ | 2 | $-\sqrt{2}$ | 0 | 0 | 0 | -2 | $\sqrt{2}$ | 0 | 0 | 0 | | |

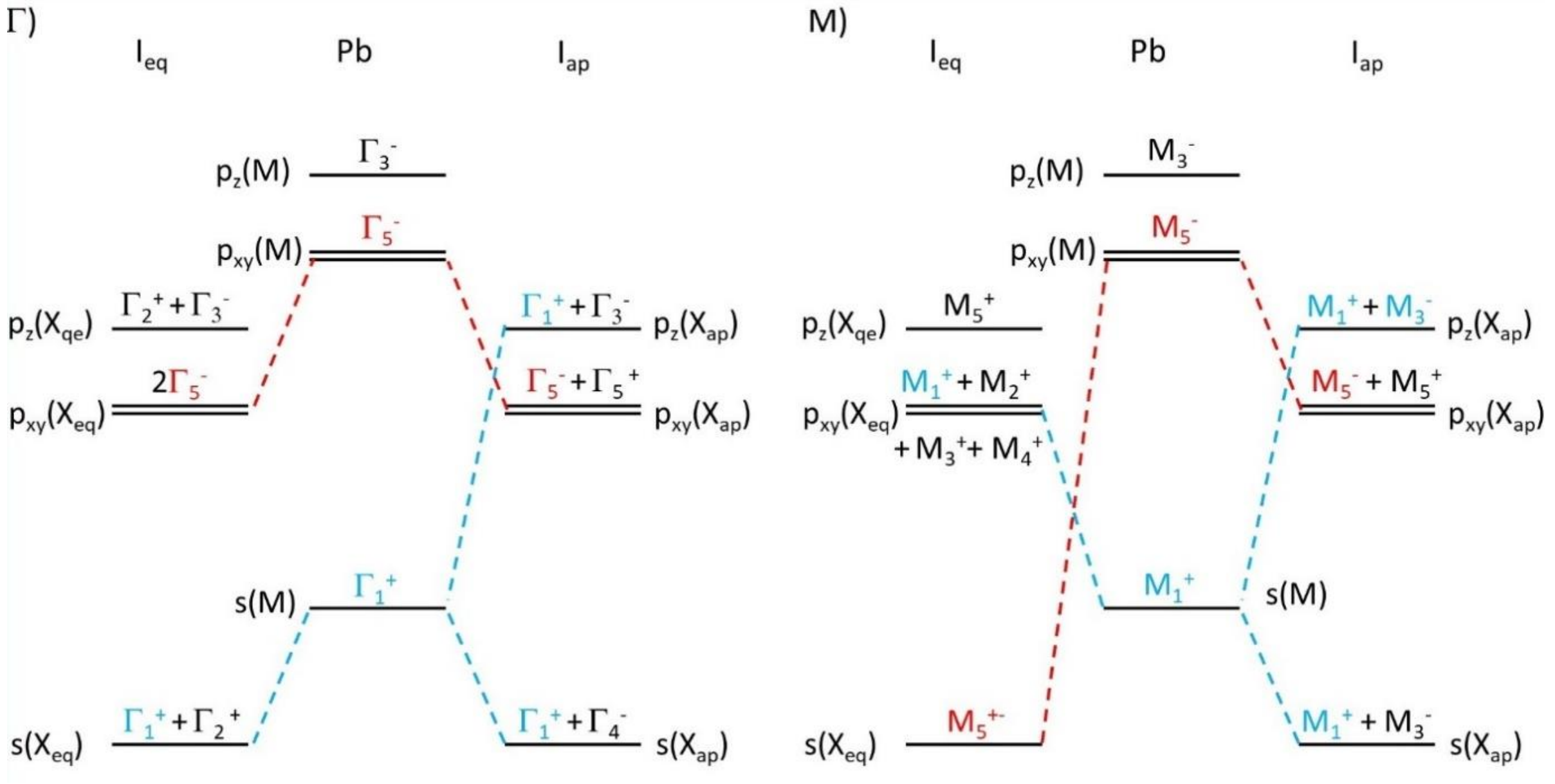

**Figure SI1.** Symmetry-analysis Symmetry adapted linear Combination of Atomic Orbitals for minimal model, $Cs_2PbX_4$, with space group symmetry P4mmm (see Figure 1a). Left and right panels refer to the atomic hybridization scheme at the Γ and M points of the Brillouin zone, respectively. Irreducible representations of Valence and Conduction Band Edges are indicated in blue and red, respectively.

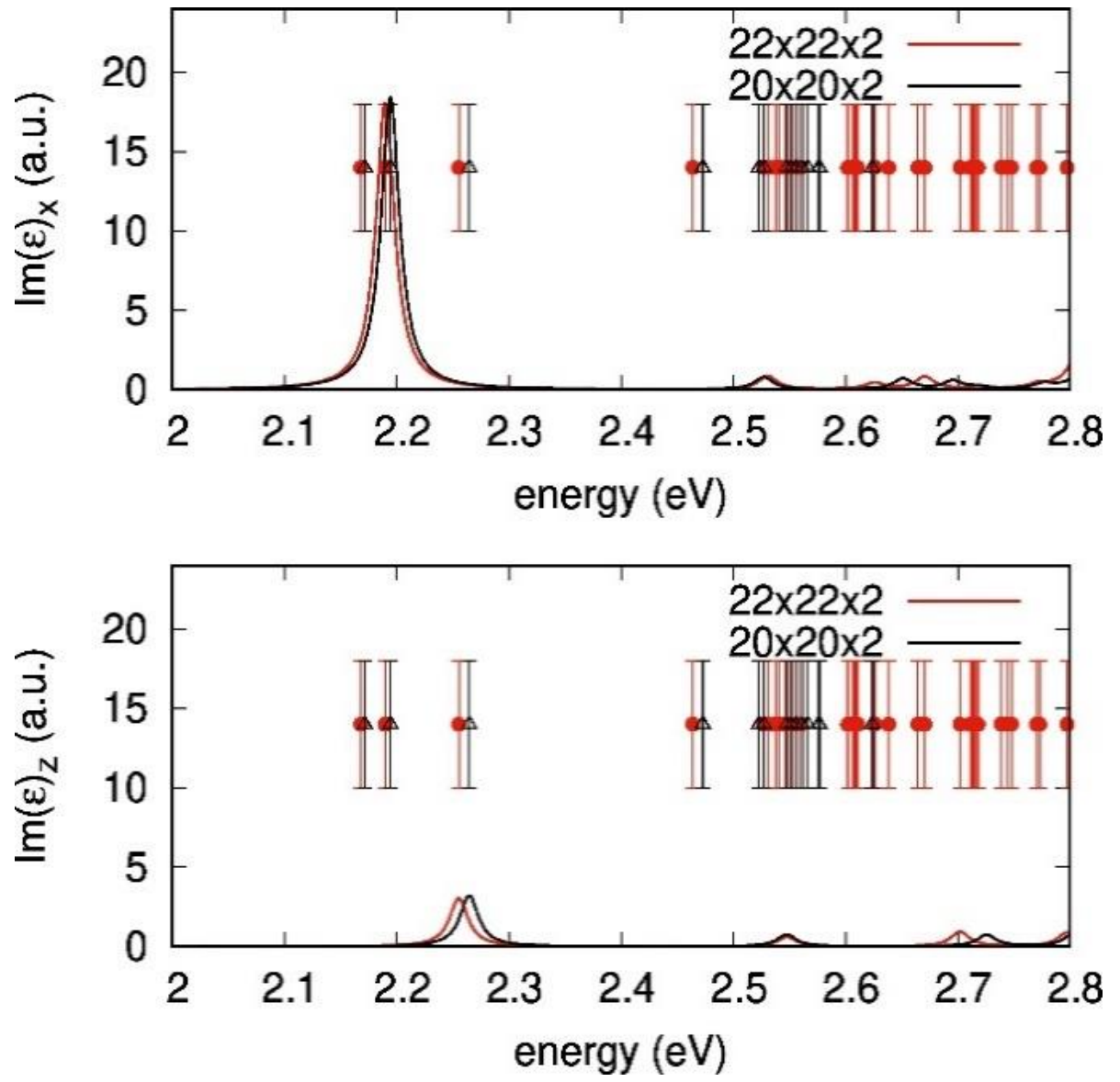

**Figure SI2.** Dependence of the BSE results with respect to the k-point mesh for the P4mmm minimal model. **Upper panel)** imaginary part of the dielectric function in the inorganic plane ($\varepsilon_x$), computed adopting 20x20x2 and 22x22x2 automatic k-point sampling. **Bottom panel)** same analysis for dielectric function along the plane stacking direction ($\varepsilon_z$). In both panels, the individual excitation energies are indicated by vertical bars, to highlight the presence of dark states.

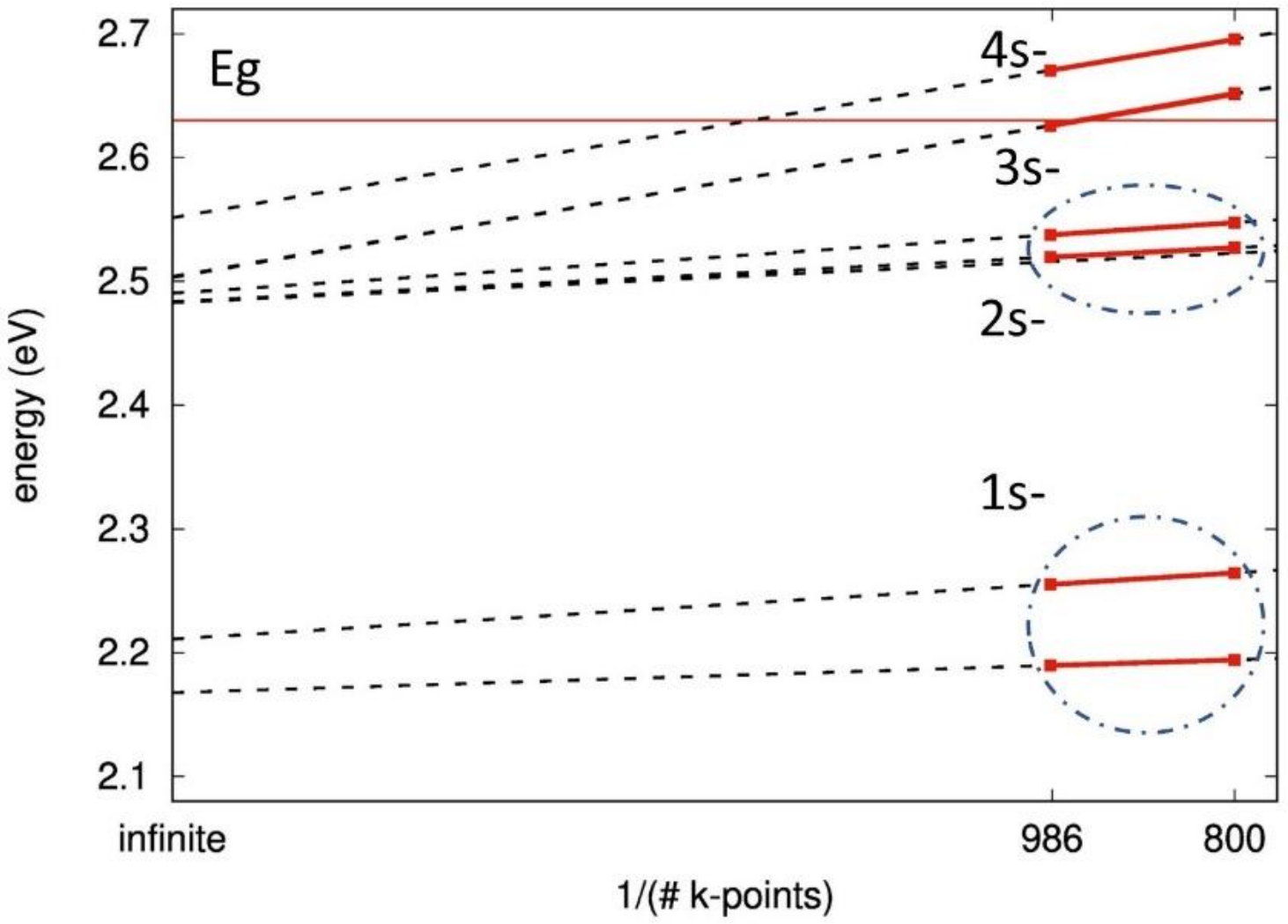

**Figure SI3.** Energetics of the bright Ns- (N=1,4) states computed from ab-initio BSE calculation for the P4mmm minimal model, plotted against the inverse of the number of k-points considered in the calculation. 800 and 986 refer to an automatic 20x20x2 and 22x22x2 sampling of the reciprocal space. Dotted lines indicate the extrapolation of the energetics for the ideal case of infinitely dense k-point mesh.

**Table SI2.** BSE solutions for the minimal model in Figure 1a. Energetics, in plane ($\mu_x$) and out-of-plane ($\mu_z$) transition dipole, proposed assignment in terms of excitonic Wannier-series components and irreducible representations at the M point of P4mmm (when possible).

| **Solution #** | **Energy (eV)** | **$\mu_x$ (norm.)** | **$\mu_z$** | **Wannier** | **IR. REP.** |
|---|---|---|---|---|---|
| 1 | 2.1715 | 0.00 | 0.00 | 1s- | $A_{1u}$ |
| 2 | 2.1942 | 0.98 | 0.00 | | $E_u$ |
| 3 | 2.1942 | 1.00 | 0.00 | | |
| 4 | 2.2645 | 0.00 | 1.00 | | $A_{2u}$ |
| 5 | 2.4726 | 0.00 | 0.00 | ? | |
| 6 | 2.4726 | 0.00 | 0.00 | | |
| 7 | 2.4729 | 0.00 | 0.00 | | |
| 8 | 2.4730 | 0.00 | 0.00 | | |
| 9 | 2.5228 | 0.00(01) | 0.00 | 2s- | $A_{1u}$ |
| 10 | 2.5272 | 0.03(95) | 0.00 | | $E_u$ |
| 11 | 2.5272 | 0.04(38) | 0.00 | | |
| 12 | 2.5474 | 0.00 | 0.23 | | $A_{2u}$ |
| 13 | 2.5516 | 0.00 | 0.00 | 2p- ($\chi_{env}=E_u$) | |
| 14 | 2.5555 | 0.00 | 0.00 | | |
| 15 | 2.5613 | 0.00 | 0.00 | | $E_g$ |
| 16 | 2.5613 | 0.00 | 0.00 | | |
| 17 | 2.5661 | 0.00 | 0.00 | | $E_g$ |
| 18 | 2.5661 | 0.00 | 0.00 | | |
| 19 | 2.5760 | 0.00 | 0.00 | | |
| 20 | 2.5767 | 0.00 | 0.00 | | |
| 21 | 2.6238 | 0.00 | 0.00 | | |
| 22 | 2.6239 | 0.00 | 0.00 | 2p- | |
| 23 | 2.6240 | 0.00 | 0.00 | ($\chi_{env}$=$A_{2u}$) | $E_g$ |
| 24 | 2.6240 | 0.00 | 0.00 | | |

Continues in the next page

| **Solution #** | **Energy (eV)** | **$\mu_x$ (norm.)** | **$\mu_x$** | **Wannier** | **IR. REP.** |
|---|---|---|---|---|---|
| 25 | 2.6372 | 0.00 | 0.00 | | |
| 26 | 2.6372 | 0.00 | 0.00 | | |
| 27 | 2.6390 | 0.00 | 0.00 | | |
| 28 | 2.6390 | 0.00 | 0.00 | 3p- | |
| 29 | 2.6397 | 0.00 | 0.00 | ($\chi_{env}$=$E_u$) | |
| 30 | 2.6397 | 0.00 | 0.00 | | |
| 31 | 2.6415 | 0.00 | 0.00 | | |
| 32 | 2.6415 | 0.00 | 0.00 | | |
| 33 | 2.6513 | 0.00 | 0.00 | | $A_{2u}$/$A_{1u}$? |
| 34 | 2.6513 | 0.00(01) | 0.00 | 3s- | |
| 35 | 2.6517 | 0.04(34) | 0.00 | | $E_u$ |
| 36 | 2.6518 | 0.02(76) | 0.00 | | |
| 37 | 2.6695 | 0.00 | 0.00 | 3p- | |
| 38 | 2.6695 | 0.00 | 0.00 | ($\chi_{env}$=$A_{2u}$) | |
| 39 | 2.6696 | 0.00 | 0.00 | | |
| 40 | 2.6696 | 0.00 | 0.00 | | |
| 41 | 2.6896 | 0.00 | 0.00 | | $A_{1u}$ |
| 42 | 2.6955 | 0.02(76) | 0.00 | 4s- | $E_u$ |
| 43 | 2.6955 | 0.02(77) | 0.00 | | |
| 44 | 2.7182 | 0.00 | 0.00(01) | | $A_{2u}$ |

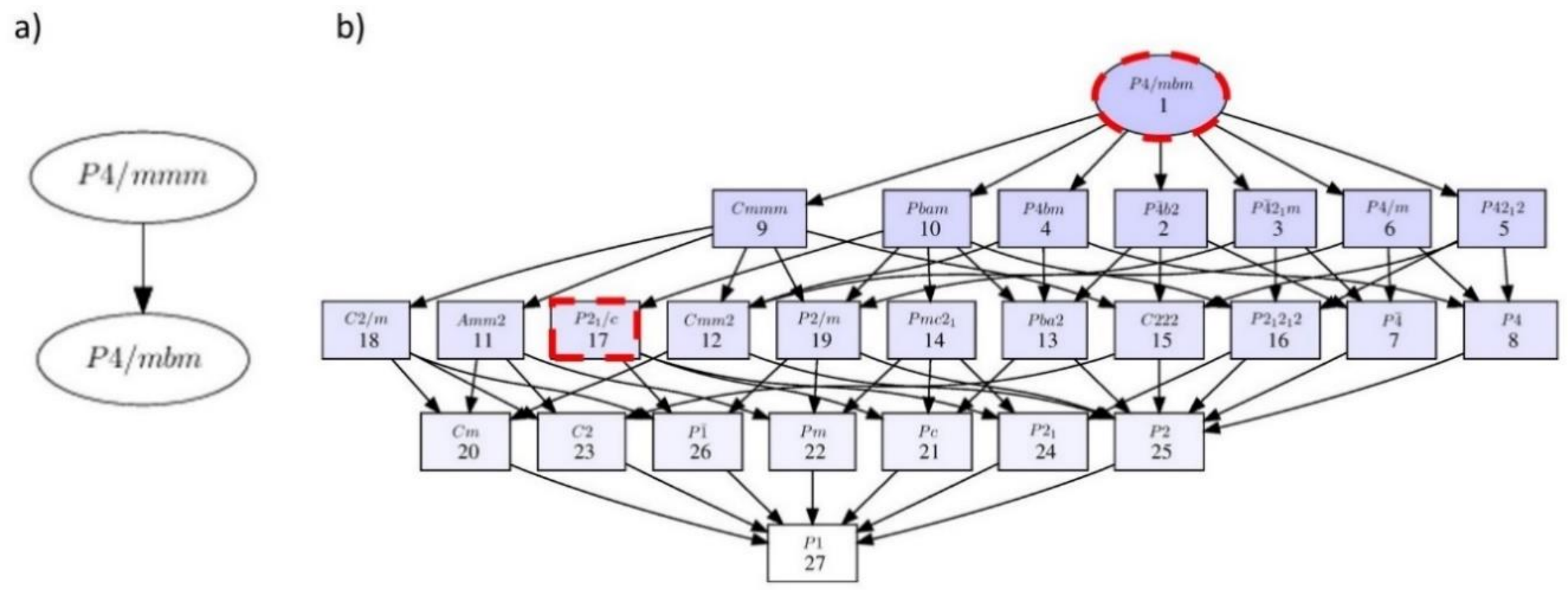

**Figure SI4.** Space group determination for layered perovskite model containing octahedral tilting in Figure 3a; **a)** space group relation from the minimal to the double model, as obtained from the program SUBGROUPGRAPH from the Bilbao Crystallographic Server; **b)** group-subgroup relations from the double to the distorted model, as obtained from the SUBGROUP program of the Bilbao Crystallographic Server.

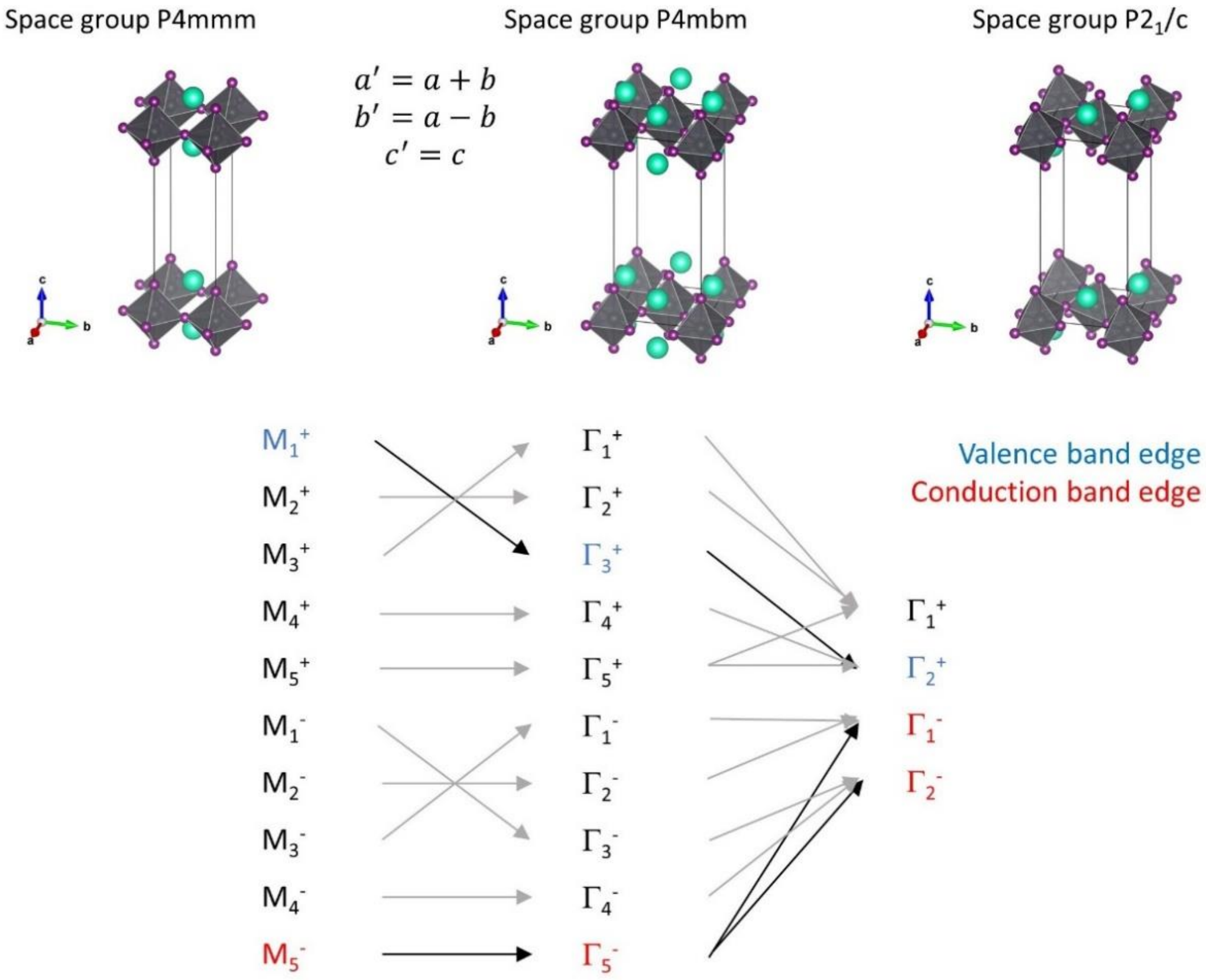

**Figure SI5.** Subduction problem related to the folding of the Brillouin zone from P4mmm to P4mbm when doubling the unit cell, and from P4mbm to P2$_1$/c when introducing anti-ferrodistorsive octahedral tilting. Irreducible representations of the valence and conduction band edges, as obtained from the CORREL program of the Bilbao Crystallographic Server, are indicated.

**Table SI3.** Character table of $C_{2h}$ point group and $P2_1/c$ space group at Γ.

| $C_{2h}$ | P21c | E | $C_2$ | i | $\sigma_h$ | | |
|---|---|---|---|---|---|---|---|
| | | | simple-group | | | | |
| $A_g$ | $\Gamma_1^+$ | 1 | 1 | 1 | 1 | | $x^2$,$y^2$,$z^2$,xy |
| $B_g$ | $\Gamma_2^+$ | 1 | -1 | 1 | -1 | | xz,yz |
| $A_u$ | $\Gamma_1^-$ | 1 | 1 | -1 | -1 | z | |
| $B_u$ | $\Gamma_2^-$ | 1 | -1 | -1 | 1 | x,y | |
| | | | double-group | | | | |
| $^1E_{1/2g}$ | $\Gamma_3^+$ | 1 | i | 1 | i | | |
| $^2E_{1/2g}$ | $\Gamma_4^+$ | 1 | -i | 1 | -i | | |
| $^1E_{1/2u}$ | $\Gamma_3^-$ | 1 | i | -1 | -i | | |
| $^2E_{1/2u}$ | $\Gamma_4^-$ | 1 | -i | -1 | i | | |

**Table SI4.** Character table of point group $C_{4v}$ and of the space group P4mm at the M.

| $C_{4v}$ | P4mm<br>M | E | $2C_4(z)$ | $C_2$ | $2\sigma_v$ | $2\sigma_d$ | | |
|---|---|---|---|---|---|---|---|---|
| | | | Simple-group | | | | | |
| $A_1$ | $M_1$ | 1 | 1 | 1 | 1 | 1 | | $x^2+y^2$, $z^2$ |
| $A_2$ | $M_4$ | 1 | 1 | 1 | -1 | -1 | z | |
| $B_1$ | $M_2$ | 1 | -1 | 1 | 1 | -1 | | $x^2-y^2$ |
| $B_2$ | $M_3$ | 1 | -1 | 1 | -1 | 1 | | xy |
| E | $M_5$ | 2 | 0 | 2 | 0 | 0 | x,y | xz, yz |
| | | | double-group | | | | | |
| $E_{1/2}$ | $M_6$ | 2 | $\sqrt{2}$ | 0 | 0 | 0 | | |
| $E_{3/2}$ | $M_7$ | 2 | $-\sqrt{2}$ | 0 | 0 | 0 | | |

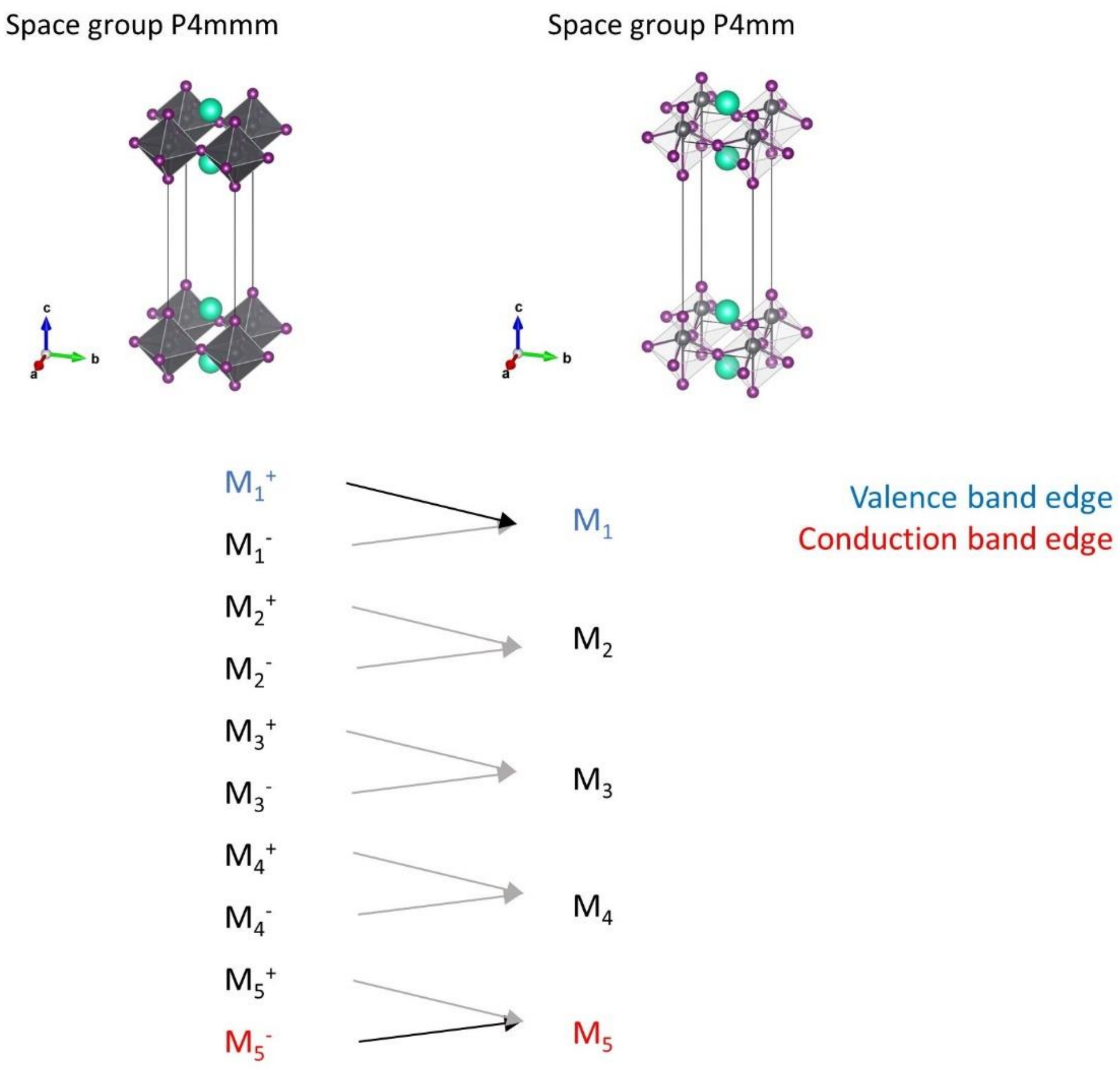

**Figure SI6.** Subduction problem from P4mmm to P4mm when introducing (out-of-plane) polar distortions and related to the irreducible representations of the valence and conduction band edges, as obtained from the CORREL program of the Bilbao Crystallographic Server.

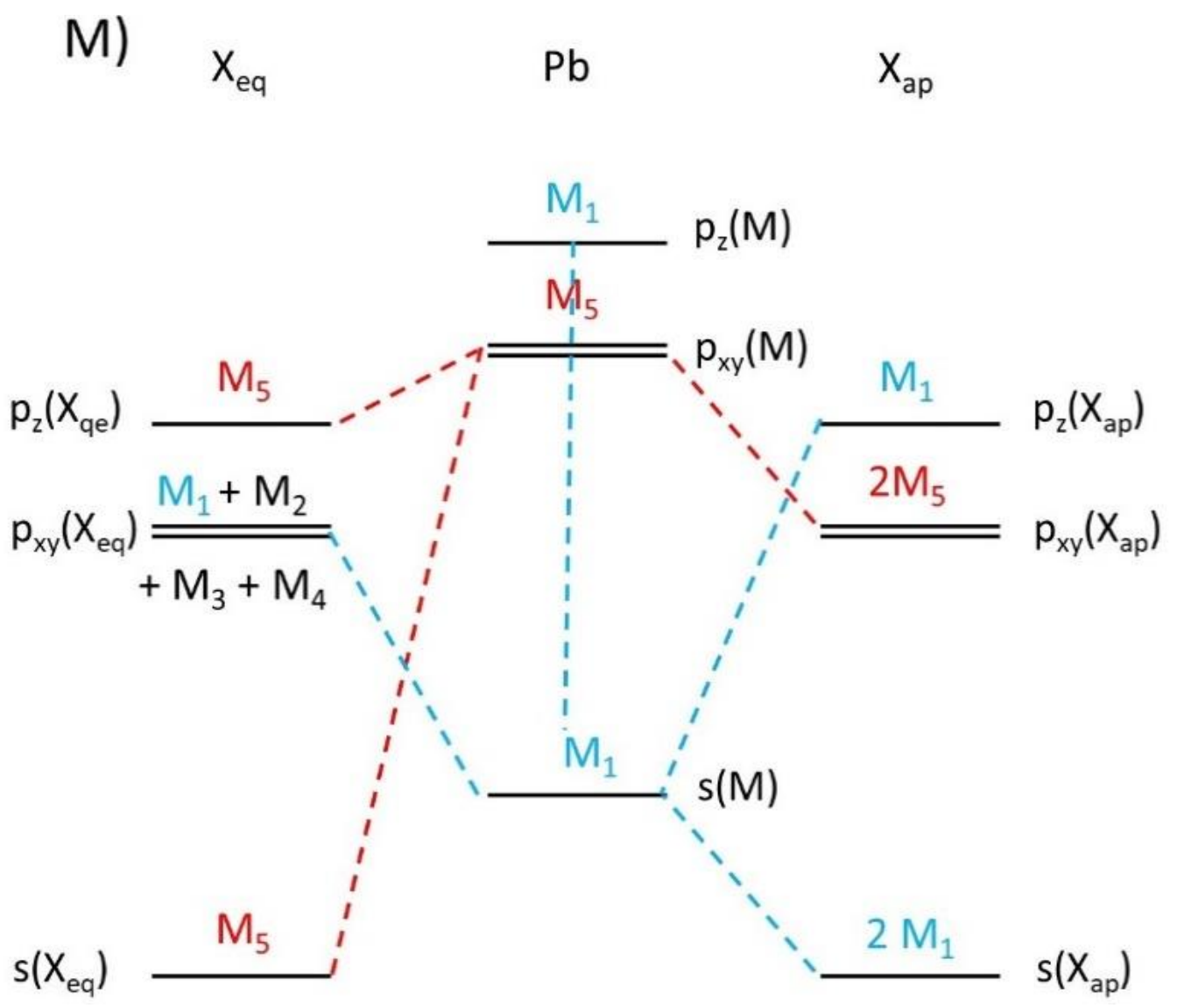

**Figure SI7.** Symmetry adapted linear Combination of Atomic Orbitals for $Cs_2PbX_4$ layered halide perovskite model with space group symmetry P4mm at the M point of the corresponding Brillouin zone.

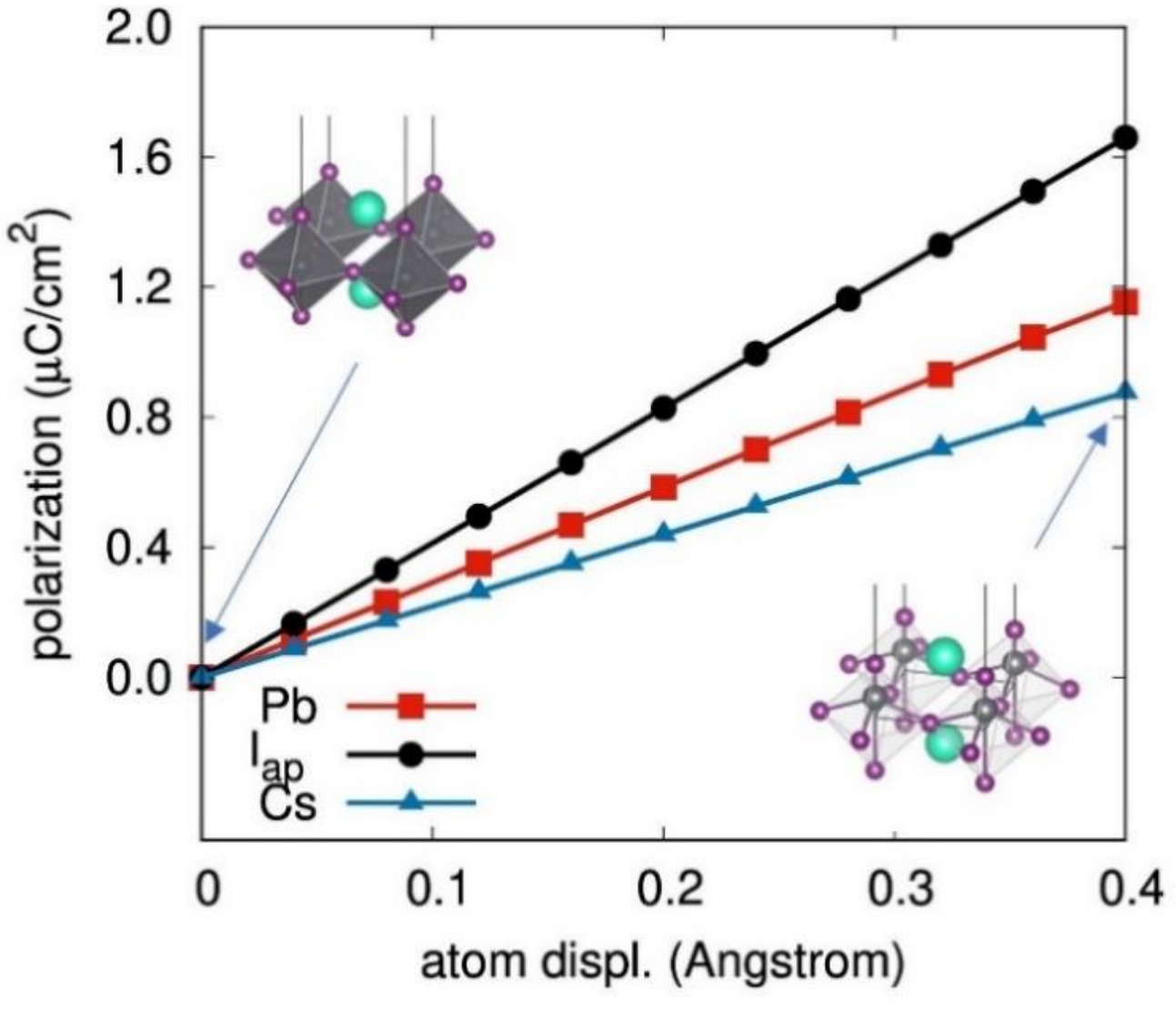

**Figure SI8.** Berry phase calculation of the spontaneous polarization of Pb, $I_{ap}$, and Cs polar-distorted models. Non-polarized reference structure corresponds to the minimal $Cs_2PbI_4$ model with P4mmm space group symmetry.

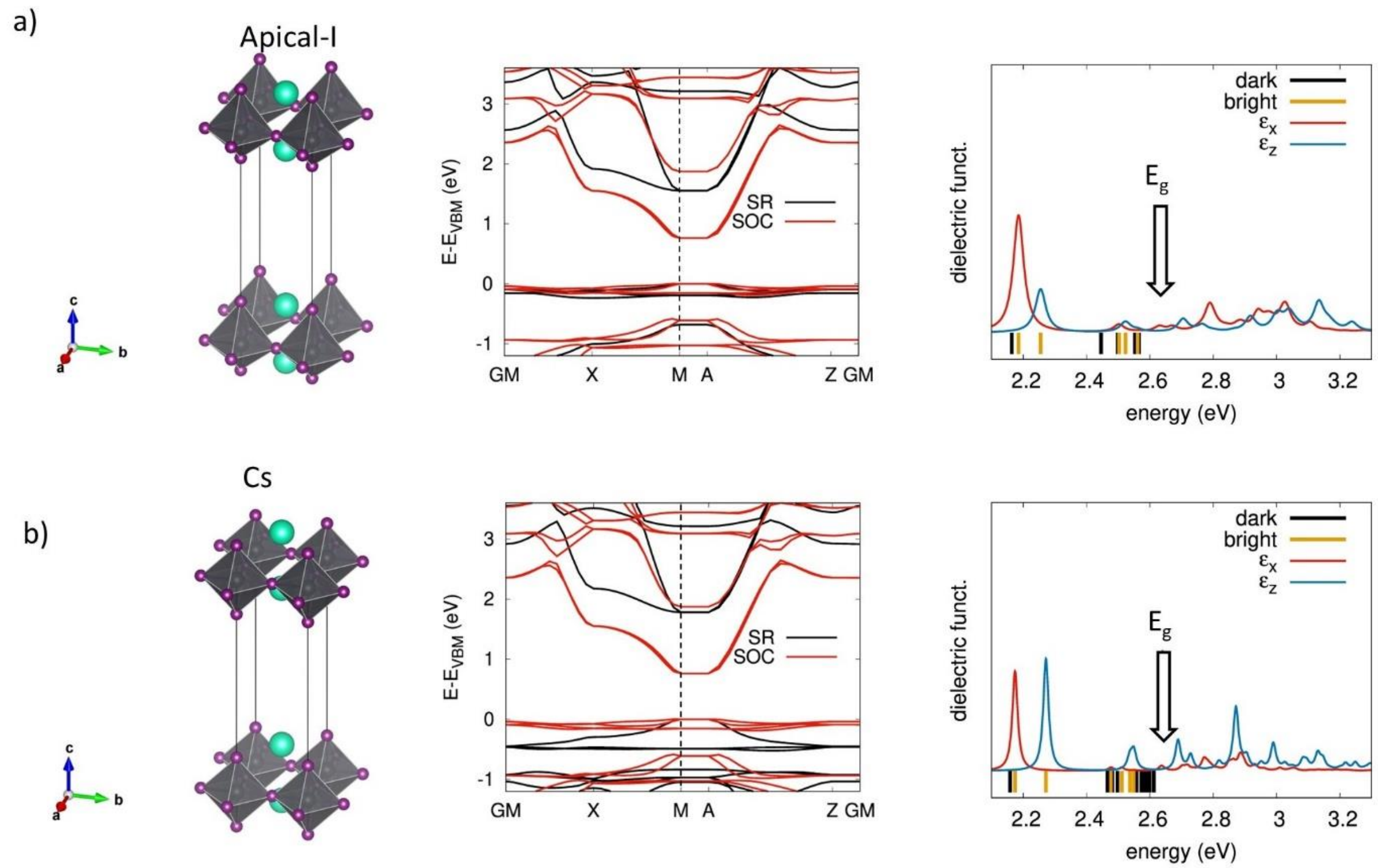

**Figure SI9.** Results of DFT (PBE without and with SOC) and BSE calculations performed on the $Cs_2PbI_4$ minimal model structure with out-of-plane polar displacement of **a)** I-apical and **b)** Cs atom.